\documentclass[a4paper,11pt]{article}
\pdfoutput=1 % if your are submitting a pdflatex (i.e. if you have
\usepackage{jcappub} % for details on the use of the package, please
\usepackage{tikz}
\usepackage{bm}
\newcommand*{\circled}[1]{\lower.7ex\hbox{\tikz\draw (0pt, 0pt)%
    circle (.5em) node {\makebox[1em][c]{\small #1}};}}

\usepackage{subcaption}

\title{Quasinormal modes of dark matter core-black hole spacetime}

\affiliation{College of Physics, Guizhou University, Guiyang 550025, China}
\author[1]{Min Zhao}
\author[1]{Meirong Tang}
\author[1,*]{Zhaoyi Xu}
\emailAdd{zyxu@gzu.edu.cn(Corresponding author)}

\abstract{
In the galactic core, when the scale of dark matter is small, the distribution of dark matter is that of a constant density dark matter core.Considering the case of a supermassive black hole coupled to a constant density dark matter core, we study the quasinormal modes of the black hole in the constant density dark matter core black hole system and calculate the quasinormal modes frequency of the black hole using the third order WKB approximation and the prony method.In addition, we study the effect of the constant density dark matter core parameter $r_0$ on the quasinormal modes of black holes in the vicinity of black holes.As the angular quantum number increases, the ringdown process becomes closer and closer to the case of the ringdown process of a schwarzschild black hole.The presence of a constant density dark matter core affects the quasinormal modes of the black hole, with relative deviations on the order of $10^{-15}-10^{-13}$ with respect to the detector.These features suggest that with future improvements in detector accuracy, we can use them for the detection of gravitational waves in the spacetime of constant density dark matter core-black hole systems, which in turn opens up the possibility of understanding the behavior of dark matter in the vicinity of black holes.
}

\begin{document}
\maketitle
\flushbottom

\section{Introduction}
\label{intro}

Observations in recent decades have shown that there is a large amount of dark matter in the universe (which does not participate in electromagnetic interactions), and that it accounts for 23\% of the matter in the universe \cite{Zwicky:1942zz}.Observations of the rotation curves of spiral galaxies, the cosmic microwave background radiation, and the large-scale structure of the universe can provide indirect evidence for the existence of dark matter \cite{1970ApJ:159379R,1992ApJ396L1S,Melott1983rn}.This observational evidence has led physicists to propose a large number of dark matter models, the most successful of which is the cold dark matter model.The cold dark matter model is relatively successful in explaining the large-scale structure of the universe and galaxy formation, but has some problems in explaining small-scale structure \cite{Bullock:2017xww}.For example, the cold dark matter model(CDM) gives a dark matter density distribution in the cores of spiral galaxies that satisfies the $1/r$ form \cite{Navarro:1995iw}.However, in 2001 de Block et al. used observations of the rotation curves of low surface brightness galaxies to find that the dark matter distribution in the core of galaxies is a dark matter core \cite{2002AA...385..816D}.It is the core/cusp problem of the cold dark matter model.Against this background, physicists have proposed self-interacting dark matter models, warm dark matter models and Fuzzy dark matter models, etc., which alleviate the core/cusp problem, etc. to some extent.The constant density core of dark matter in the centers of galaxies plays an important role in understanding the small-scale effects of dark matter, and by measuring the constant density core of dark matter, it is possible to place some constraints on the dark matter model \cite{Rocha:2012jg,Bode:2000gq,Maccio:2012qf}.On the other hand, the interaction of dark matter with supermassive black hole in the center of galaxies is a very interesting problem, and how to construct a spacetime metric for the dark matter constant density core-black hole system from the proper dark matter equation of state is the key to study this problem.In 2021 Gong et al. obtained a numerical solution for the dark matter constant density core-black hole system by making reasonable assumptions about the equation of state of dark matter \cite{Gong:2020lev}.Recently, with the successful detection of stochastic gravitational wave backgrounds, it has become possible to understand observationally the supermassive black hole-dark matter interactions at the core of galaxies.

In the final stage of black hole merger, the change of waveforms of gravitational waves is correlated with the quasinormal modes(QNMs) of the black hole, and by studying the quasinormal modes of the black hole, the fundamental nature of the black hole can be understood indirectly.In general, the frequency of the quasinormal modes of a black hole can be described by a complex number, where the real part describes the speed of oscillation and the imaginary part describes the speed of decay.The evolution of the black hole perturbation can be divided into three phases, namely the initial wave burst phase, the oscillatory decay phase, and the power-law trailing phase.The second of these phases, the oscillatory decay phase, is the quasinormal modes phase, which carries information about black holes and is an important way for us to study black holes\cite{Regge:1957td,Zerilli:1970se,Vishveshwara:1970zz,PhysRevD73124040,LpezOrtega2006,Liu:2007zze,Cho:2003qe,Jing2004,PhysRevD.71.024007,PhysRevD69084015}.On the other hand physicists have carried out a lot of research on the quasinormal modes of various special black holes\cite{Berti:2009kk,Toshmatov:2017bpx,Aneesh:2018hlp,Gogoi:2023kjt,Churilova:2019cyt,Kanzi:2021cbg,Vishvakarma:2023csw,Abbas:2023pug,Zerilli:1974ai,Teukolsky:1972my,PhysRevD.64.084017,Liu:2021xfb,Liu:2022ygf,Yang:2021cvh,Yang:2022xxh,Daghigh:2022pcr,Zhao:2023tyo,Cardoso:2021wlq,Figueiredo:2023gas,Konoplya:2022hbl,Jusufi:2022jxu}, and in two recent papers\cite{Daghigh:2022pcr,Zhao:2023tyo} they have investigated the coupling of black holes and dark matter based on the fact that dark matter around black holes is a distribution of spikes, and they have investigated the detection of the existence of dark matter in the vicinity of supermassive black holes by means of gravitational wave detections emitted in the ringdown process of the black hole perturbation.And here there exists an opposite distribution of dark matter, i.e., a situation where the dark matter in the vicinity of a black hole is a distribution of constant density dark matter cores, which we have studied accordingly based on such a situation\cite{Gong:2020lev,KuziodeNaray2006,KuziodeNaray2010}.However, as mentioned before, the dark matter in the cores of galaxies presents a constant density dark matter core image, so considering a quasinormal modes  for constant density dark matter core-black hole systems would be more consistent with the observations.In the first paragraph we mentioned the existence of a system of supermassive black holes with constant density dark matter cores in the galactic core and obtained the spacetime gauge of this system by solving the einstein field equations.This opens up the possibility of studying constant density dark matter cores from a quasinormal modes pathway.In this work, we will compute the quasinormal modes of the constant density dark matter core-black hole system based on the spatio-temporal metrics of the system obtained by Gong et al. In this way, we will be able to understand the various properties of the constant density dark matter core through the quasinormal modes pathway\cite{Gong:2020lev}.

The paper is organized as follows. In section 2 the spacetime line elements of the constant density dark matter core-black hole system are presented. The effective potentials for the scalar and gravitational field perturbation scenarios are computed in Section 3. In section 4 the methodology for calculating the QNM frequencies is presented. Quasinormal modes for various scenarios are calculated in Section 5. Section 6 Summary.

\section{Supermassive black hole in the dark matter constant density core}
\label{metric}
Numerous observations have shown that dark matter in the cores of galaxies exhibits a constant density core structure, and that there is generally at least one supermassive black hole in the core of galaxies.Therefore, the space-time line element corresponding to the dark matter constant density core-black hole system is the key to study the interaction between dark matter constant density core and black hole.In the literature Gong et al. the authors have been able to obtain a spatio-temporal metric describing the constant density core-black hole system \cite{Gong:2020lev}.

The interaction of a dark matter constant density core with a black hole evolves in time, but in the approximate case one can consider the system as a static spherically symmetric metric with the following spacetime line element,

\begin{equation}\label{Z} 
ds^2=e^Adt^2-e^Bdr^2-r^2(d\theta^2+\sin^2\theta d\varphi^2),
\end{equation}

This space-time line element satisfies the Einstein field equation when considering the constant density core of dark matter.The energy momentum tensor is as follows $T_\nu ^\mu = diag[\rho,-p,-p,-p]$,where $\rho$ is the energy density of the dark matter and $p$ is the pressure of the dark matter.Under all the same conditions, the Einstein field equations for this system can be simplified as

\begin{equation}
N_x^{\prime}+[\frac{F_x^{\prime}}{2(1+F)}-2] N+\frac{N^2}{2}+\frac{F_x^{\prime}-2F}{1+F} = 0.
\end{equation}

According to the assumptions made in the literature Gong et al., the equation of state can be written in the following form

\begin{equation}
p=\frac{(1+F)N+F}{8\pi r^2}\approx 2\epsilon V_{rot}^2 \rho+\frac{\gamma - \epsilon - 1}{2\pi}(\frac{V_{rot}^2}{r})^2,
\end{equation}

The relationship between the unknown functions $N$, $F$, $x$ and the space-time metric coefficients are $N=A_x^{\prime} ,F=e^{-B}-1 ,x=ln(r)$,$V_{rot}$is the particle's rotation velocity, $\gamma$ and $\epsilon$ are constants.

The literature Gong et al. gives the expression of the metric coefficient, its mathematical form is

\begin{equation}
\frac{1}{g(r)}=e^B=\frac{1}{1-\frac{b}{r}\sqrt{1+(\frac{r}{r_0})^{\frac{8\epsilon-1}{2\epsilon}}}},
\end{equation}

\begin{equation}
f(r)=e^A=(1-\frac{b}{r})k(r),
\end{equation}

thus
\begin{equation}\label{Z1}
ds^2=(1-\frac{b}{r})dt^2-\frac{1}{1-\frac{b}{r}\sqrt{1+(\frac{r}{r_0})^3}}dr^2-r^2(d\theta^2+\sin^2\theta d\varphi^2).
\end{equation}

It is calculated that the function $k(r)$ tends to 1 when r takes a large value.Thus $k(r) = 1$ is a reasonable approximation for the constant density core of dark matter.The parameters $b = 2M$ and $r_0$ is characteristic radius of the constant density of dark matter,and M is the black hole mass.As $r_0\to\infty$, the spacetime metric degenerates to the Schwarzschild black hole form.When the parameter $\epsilon = 0.5$, the spacetime metric (\ref{Z}) describes a typical dark matter constant density core-black hole system.

Some dark matter related parameters are given here\cite{Gong:2020lev,KuziodeNaray2006,KuziodeNaray2010}, and the low surface brightness galaxy F568-3 is chosen as the research object. We know that the parameters of the constant density dark matter core are $r_0$= 2.83 kpc and the density of the constant density dark matter core is $\rho_0 = (40 \pm 6) \times 10^{-3} M_{\odot} pc^{-3}$. The total mass of the galaxy-halo system with a constant density dark matter core is $M_{tot} = 28 \times 10^{10} M_{\odot}$.The mass of the black hole at the center of a low surface brightness galaxy was given in a recent paper\cite{2016MNRAS4553148S}-a mass of $M = 5.62\times10^6M_{\odot}$.For the data given above, we will perform a unit conversion to the constant density dark matter core parameter $r_0=2.83kpc/(2GM/c^2)$.

\section{Perturbative modeling of the dark matter constant density core-black hole.}
\label{cdm-pro}
\subsection{Scalar field perturbation model for dark matter constant density core-black hole}
The Klein-Gordon equation in the background of curved spacetime is as follows

\begin{equation}\label{Z2}
\frac{1}{\sqrt{-g}}\partial_\mu(\sqrt{-g}g^{\mu\nu}\partial_\nu\psi)=0,
\end{equation}

by the separated variables method, the wave function $\psi$ can be written as
\begin{equation}
\psi=R(r)Y_{lm}(\theta,\varphi)e^{-i\omega t}/r.
\end{equation}

Next, a coordinate transformation is applied to the KG equation. Its corresponding the tortoise coordinate can be expressed as
\begin{equation}
dr_*=\frac{1}{\sqrt{f(r)g(r)}}dr,
\end{equation}

$Y_{lm}(\theta,\varphi)$ in the wave function $\psi$ is the spherical harmonic function, $l$ is the angular quantum number, $m$ is the magnetic quantum number.We substitute the background metric Equation (\ref{Z1}) into Equation (\ref{Z2}), so the KG equation will be transformed into the following form of Schrödinger's equation

\begin{equation}\label{L}
\frac{d^2\psi}{dr_*^2}+(\omega^2-V(r))\psi=0,
\end{equation}

the potential function $V(r)$ is expressed as
\begin{equation}\label{A}
V(r)=(1-\frac{b}{r})\bigg[\frac{b(1-\frac{b}{r}\sqrt{1+(\frac{r}{r_0})^3})}{2r^3(1-\frac{b}{r})}-\frac{3b}{4r_0^3\sqrt{1+(\frac{r}{r_0})^3}}+\frac{b\sqrt{1+(\frac{r}{r_0})^3}}{2r^3}+\frac{l(l+1)}{r^2}\bigg],
\end{equation}

when the dark matter constant density core disappears, the potential function of the scalar field degenerates to the Schwarzschild black hole case
\begin{equation}\label{B}
V(r)=(1-\frac{b}{r})(\frac{l(l+1)}{r^2}+\frac{b}{r^3}).
\end{equation}
The figures of equations (\ref{A}) and (\ref{B}) are shown in Part 5.

\subsection{Gravitational field perturbation model for dark matter constant density core-black hole}
\label{uldm-pro}
According to the gravitational field perturbation theory of black holes, the perturbation metric can be decomposed into the sum of the background metric and the small perturbations, i.e. $g_{\mu\nu}=\overline{g}_{\mu\nu}+h_{\mu\nu}$,where $h_{\mu\nu}$ is a samll quantity and $g_{\mu\nu}$ is the background metric \cite{Regge:1957td}.

For this metric perturbation, the Christoffel symbol can be expressed as
\begin{equation}\label{AZZ1}
\Gamma_{\mu\nu}^{\alpha} = \overline{\Gamma}_{\mu\nu}^{\alpha}+\delta \Gamma_{\mu\nu}^{\alpha},
\end{equation}

here the expression for $\delta\Gamma_{\mu\nu}^{\alpha}$ is as follows
\begin{equation}\label{AZZ2}
\delta\Gamma_{\mu\nu}^{\alpha}=\frac{1}{2}g^{\alpha\lambda}(h_{\mu\lambda;\nu}+h_{\nu\lambda;\mu-h_{\mu\nu;\lambda}}).
\end{equation}

In the space-time background of the dark matter constant density core-black hole system, we use the odd perturbation form of the Regge and Wheeler canonical transformation with the expression
\begin{equation}\label{AZZ5}
h_{\mu\nu}=
\left(\begin{array}{cccc}
0&0&0&h_0(t,r)\\
0&0&0&h_1(t,r)\\
0&0&0&0\\
h_0(t,r)&h_1(t,r)&0&0
\end{array}
\right)\sin\theta\partial\theta P_l(\cos\theta),
\end{equation}

$P_l(\cos\theta)$ is a Legendre polynomial,where we set $y_P(\theta)=\sin\theta \partial\theta P_l(\cos\theta)$, and we have
\begin{equation}\label{AZZ6}
y^{\prime\prime}_P(\theta)-\cot(\theta)y_P^{\prime}(\theta)= -l(l+1)y_P(\theta).
\end{equation}

Einstein's field equations
\begin{equation}
E_{\mu\nu}=R_{\mu\nu}-\frac{1}{2}Rg_{\mu\nu}=8\pi G_NT_{\mu\nu}
\end{equation}

$R_{\mu\nu}$ in the above equation is the Ricci tensor, $R$ is the Ricci scalar, and the presence of dark matter produces the energy-motion tensor $T_{\mu\nu}$. Based on the literature \cite{Zhao:2023tyo,Zhang:2021bdr}, we ignore the perturbation of dark matter, and will get $E_{13}$ and $E_{23}$ as respectively

\begin{equation}\label{QW}
E_{13}=\frac{\partial^2 h_{1}}{\partial t^2}-\frac{\partial^2 h_{0}}{\partial r \partial t}+\frac{2}{r}\frac{\partial h_{0}}{\partial t}+\frac{f(r)[l(l+1)-2g(r)-rg^{\prime}(r)]}{r^2}h_{1}-\frac{g(r)f^{\prime}(r)}{r}h_{1},
\end{equation}

\begin{equation}\label{QE}
 E_{23}=\frac{g(r)f^{\prime}(r)}{2 f(r)} h_{1} +\frac{g^{\prime}(r)}{2}h_{1}+g(r)\frac{\partial h_{1}}{\partial r}-\frac{1}{f(r)} \frac{\partial h_{0}}{\partial t},
\end{equation}

Definition $\psi = \frac{\sqrt{f(r)g(r)}}{r}h_1(t,r)$,the tortoise coordinate $dr_*=\frac{1}{\sqrt{f(r)g(r)}}dr$.By using the above equation we get

\begin{eqnarray}\label{M}
&&\frac{\partial \psi^2}{\partial t^2}-\frac{\sqrt{f(r)g(r)}}{r}\frac{\partial}{\partial r}\bigg[\frac{1}{2}\frac{\partial}{\partial r}(r\psi\sqrt{f(r)g(r)})+\frac{f(r)g(r)}{2}\frac{\partial}{\partial r}(\frac{r\psi}{\sqrt{f(r)g(r)}})\bigg]+\frac{2\sqrt{f(r)g(r)}}{r^2}\bigg[\frac{1}{2}\frac{\partial}{\partial r}(r\psi\sqrt{f(r)g(r)})\nonumber\\
&&+\frac{f(r)g(r)}{2}\frac{\partial}{\partial r}(\frac{r\psi}{\sqrt{f(r)g(r)}})\bigg]+\frac{f(r)[l(l+1)-2g(r)-rg^{\prime}(r)]}{r^2}\psi-\frac{g(r)f^{\prime}(r)}{r}\psi=0. 
\end{eqnarray}

Reduce equation (\ref{M}) to a form similar to Schrödinger's equation, e.g.:
\begin{equation}\label{Y}
\frac{d^2\psi}{dt^2}-\frac{d^2\psi}{dr_*^2}+V(r)\psi=0,
\end{equation}

Substituting the $f(r)$, $g(r)$ function yields the gravitational perturbation potential function as
\begin{equation}\label{C}
V(r)=(1-\frac{b}{r})\bigg[\frac{l(l+1)}{r^2}+\frac{9b}{4r_0^3\sqrt{1+(\frac{r}{r_0})^3}}-\frac{3b(1-\frac{b}{r}\sqrt{1+(\frac{r}{r_0})^3})}{2r^3(1-\frac{b}{r})}-\frac{3b\sqrt{1+(\frac{r}{r_0})^3}}{2r^3}\bigg].
\end{equation}

When the dark matter constant density core disappears, the potential function of the gravitational field perturbation degenerates to the Schwarzschild black hole situation,
\begin{equation}\label{D}
V(r)=(1-\frac{b}{r})(\frac{l(l+1)}{r^2}-\frac{3b}{r^3}).
\end{equation}
The figures of equations (\ref{C}) and (\ref{D}) are shown in Part 5.

\section{The method}

\subsection{The WKB method}
\label{test-test particle}
For quasinormal modes frequencies were solved by Schutz and Will using the WKB approximation back in 1985\cite{Wentzel:1926aor,1985ApJ...291L..33S}.In 1987, Lyer and Will extended the WKB approximation to the third order \cite{PhysRevD.35.3632}. This improved the accuracy of the calculations.In 2003, Konoplya further extended the method up to the sixth order \cite{PhysRevD.68.024018}.The approximation has been extended to the 13th order by Matyjasek and Opala in the year 2017 \cite{PhysRevD.96.024011}.The use of the 13th order WKB approximation in this work yielded values that were significantly different from those of the 3rd and 6th order WKB approximations, and therefore this accuracy was not feasible for this study.Some researchers have further explained: It is not the case that the higher the order of the WKB approximation method, the higher the accuracy \cite{Hatsuda:2019eoj}.We therefore use the third order WKB approximation to compute the quasinormal modes of the constant density dark matter core-black hole spacetime model in scalar field perturbations and gravitational field perturbations, respectively.
The 3rd order formula for the WKB approximation method is
\begin{equation}
\omega^2=[V_0+(-2V_0^{\prime\prime})^{1/2}\Lambda]-i(n+\frac{1}{2})(-2V_0^{\prime\prime})^{1/2}(1+\Omega),
\end{equation}

here
\begin{equation}
\Lambda(n)=\frac{1}{(-2V_0^{\prime\prime})^{1/2}}[\frac{1}{8}(\frac{V_0^{(4)}}{V_0^{\prime\prime}})(\frac{1}{4}+\alpha^2)-\frac{1}{288}(\frac{V_0^{\prime\prime\prime}}{V_0^{\prime\prime}})^2(7+60\alpha^2)],
\end{equation}

\begin{eqnarray}
\Omega(n)=&&\frac{1}{-2V_0^{\prime\prime}}[\frac{5}{6912}(\frac{V_0^{\prime\prime\prime}}{V_0^{\prime\prime}})^4(77+188\alpha^2)-\frac{1}{384}(\frac{V_0^{\prime\prime\prime 2}V_0^{(4)}}{V_0^{\prime\prime 3}})(51+100\alpha^2)+\frac{1}{2304}(\frac{V_0^{(4)}}{V_0^{\prime\prime}})^2\nonumber\\
&&(67+68\alpha^2)+\frac{1}{288}(\frac{V_0^{\prime\prime\prime}V_0^{(5)}}{V_0^{\prime\prime 2}})(19+28\alpha^2)-\frac{1}{288}(\frac{V_0^{(6)}}{V_0^{\prime\prime}})(5+4\alpha^2)].
\end{eqnarray}

\subsection{The finite difference method}

In this paper, the finite difference method proposed by Gundlach, Price, and Pullin \cite{PhysRevD.49.890} is used to obtain the dynamical evolution behavior of equations (\ref{L}) and (\ref{Y}).The purpose is to obtain time domain profiles.Transformations of equations (\ref{L}) and (\ref{Y}) using the coordinates $u=t-r_{*}$ and $v=t+r_{*}$ give
\begin{equation}
-4\frac{\partial^2\psi(u,v)}{\partial_u\partial_v}=V(u,v)\psi(u,v),
\end{equation}

For the above equation, it can be discretized as \cite{Moderski:2005hf}

\begin{equation}
\psi(N)=\psi(E)+\psi(W)-\psi(S)-\Delta^2\frac{V(E)\psi(E)+V(W)\psi(W)}{8}+o(\Delta^4),
\end{equation}

These grid points in the above equations correspond to $N=(u+ \Delta ,v+\Delta),E=(u+ \Delta ,v), W=(u ,v+\Delta)$ and $S=(u ,v)$, respectively.We use Gaussian wave packets
\begin{equation}
\psi(u=u_0,v)=exp[-\frac{(v-v_c)^2}{2\sigma^2}],\\
\psi(u,v=v_0)=0,
\end{equation}

Where $\sigma=3,v_c=10$.Thus we obtain the time-domain profile by the finite difference method.In addition we use the prony method to fit the signal by superimposing a damping exponent $\psi (t)\simeq\sum\limits_{i=1}^{p}C_ie^{-i\omega_it}$,and extract frequencies from it\cite{Berti:2007dg,Churilova:2019cyt}.

\section{Quasinormal modes of the constant density core-black hole}
\label{test-scalar field}
We analyze detectability in scalar and gravitationally perturbed perturbation fields using quasinormal modes of black holes to detect constant density dark matter cores as well as from a detector perspective.

\subsection{scalar field}

\begin{figure*}[t!]
        \begin{minipage}[t]{0.44\linewidth}
         \centerline{\includegraphics[width=\textwidth]{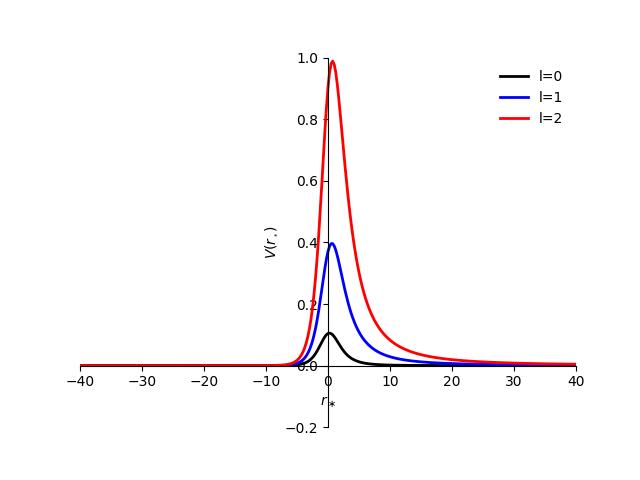}}
		\centerline{(a)}
	\end{minipage}
        \begin{minipage}[t]{0.44\linewidth}
         \centerline{\includegraphics[width=\textwidth]{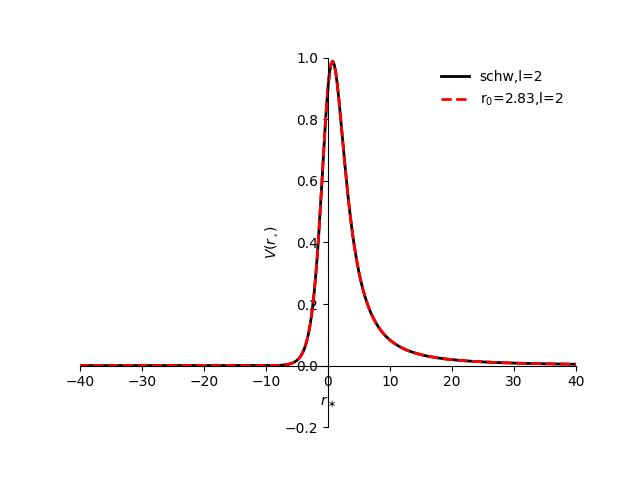}}
		\centerline{(b)}
	\end{minipage}
	\caption{Figure (a) represents the effective potential plots for $l$ = 0, 1, and 2 at a scalar field perturbation with a constant density core parameter of dark matter, $r_0$ = 2.83 kpc. Figure (b) represents the potential function plots for the comparison with the Schwarzschild black holes at a scalar field perturbation with $r_0$ = 2.83 kpc and $l$ = 2.}
	\label{fig1}
\end{figure*}

\begin{figure*}[t!]
        \begin{minipage}[t]{0.44\linewidth}
         \centerline{\includegraphics[width=\textwidth]{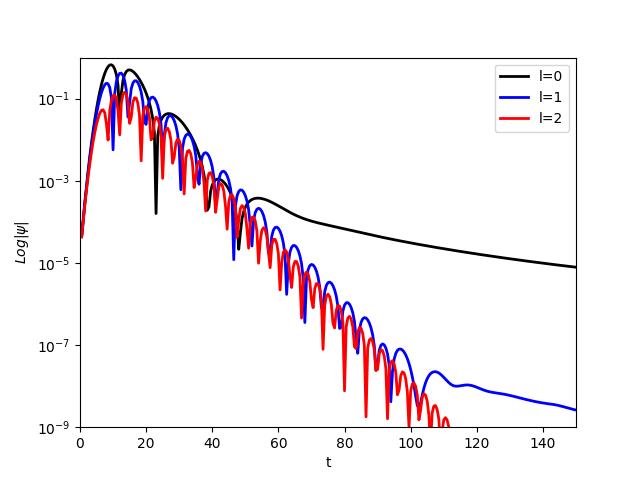}}
		\centerline{(c)}
	\end{minipage}
        \begin{minipage}[t]{0.44\linewidth}
         \centerline{\includegraphics[width=\textwidth]{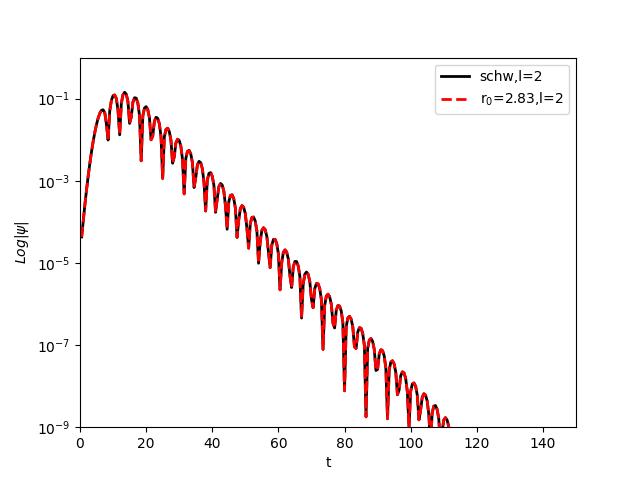}}
		\centerline{(d)}
	\end{minipage}
	\caption{Figure (c) represents the wave function plots for l = 0, 1, and 2 at a scalar field perturbation with a constant density core parameter of dark matter, $r_0$ = 2.83 kpc. (d) represents the wave function plots for the comparison with the Schwarzschild black holes at a scalar field perturbation with $r_0$ = 2.83 kpc and $l$ = 2.}
	\label{fig2}
\end{figure*}

\begin{table*}[htbp]
\caption{\label{table1}The quasinormal mode frequencies of the scalar field perturbations.}
\begin{tabular}{ccccccc}
\hline
\hline
\multicolumn{1}{c}{$l=0$}\\
\hline
\multicolumn{1}{c}{$r_0$}  &\multicolumn{1}{c}{3th ordel WKB}  &\multicolumn{1}{c}{prony method}\\
\hline
schw  &0.2092936249449866-0.23039350076033174$i$   &0.22095988563717434-0.2089322237666429$i$\\
2.83  &0.20929362495058507-0.23039350076540938$i$  &0.2209872795354999-0.20890920835877713$i$ \\
\hline
\multicolumn{1}{c}{$l=1$}\\
\hline
schw  &0.5822282327591899-0.19600272625877158$i$ &0.5871008354360577-0.1944025477566691$i$   \\
2.83  &0.5822282327591605-0.19600272625865237$i$ &0.5871008354471453-0.19440254613868468$i$ \\
\hline
\multicolumn{1}{c}{$l=2$}\\
\hline
schw  &0.9664220608737814-0.19360970973913258$i$ &0.9699450625521562-0.19172199365454512$i$  \\
2.83 &0.9664220608737852-0.19360970973914415$i$ &0.9699450625521602-0.19172199365452647$i$ \\
\hline
\hline
\end{tabular}
\end{table*}

In the scalar field, the (a) plot of Fig.\ref{fig1} we draw the image of the potential function in the spacetime of the dark matter constant density core-black hole system for angular quantum numbers $l$ of 0, 1, and 2, which are increasing with increasing angular quantum number.The (b)-plot of Fig.\ref{fig1} shows the image of the potential function for the comparison with the Schwarzschild black hole at the constant density core parameter of dark matter, $r_0$ = 2.83kpc, and the angular quantum number, $l$ = 2.The (c)-plot of Fig.\ref{fig2} is an image of the wavefunction of the dark matter constant density core-black hole system in spacetime with angular quantum numbers $l$ of 0, 1, and 2, which decays faster and faster as the angular quantum number increases.The (d)-plot in Fig.\ref{fig2} shows the wavefunction image for comparison with the Schwarzschild black hole at the constant density core parameter of dark matter, $r_0$ = 2.83, and the angular quantum number, $l$ = 2.

In table\ref{table1} we use the third order WKB approximation and prony to obtain the frequencies of the spacetime of the constant density dark matter core-black hole system in a scalar field.From table\ref{table1} we know that the presence of a constant density dark matter core influences the quasinormal modes of the black hole.The error between the frequency obtained by applying the prony method and the frequency obtained by the wkb approximation method is about 0.005-0.02, which is in good agreement.

\subsection{gravitational field}

\begin{figure*}[t!]
        \begin{minipage}[t]{0.44\linewidth}
         \centerline{\includegraphics[width=\textwidth]{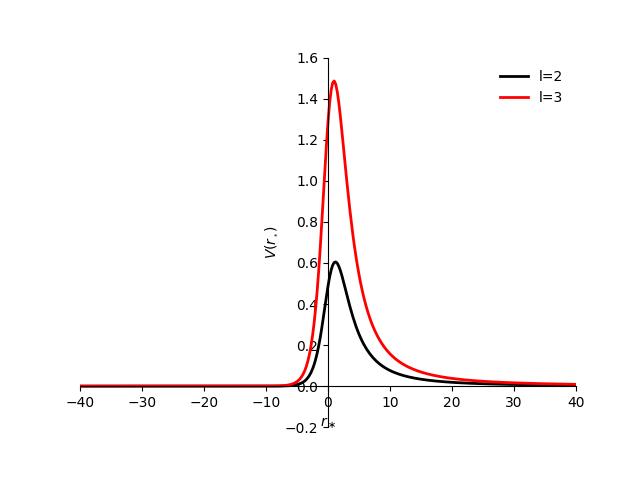}}
		\centerline{(e)}
	\end{minipage}
        \begin{minipage}[t]{0.44\linewidth}
         \centerline{\includegraphics[width=\textwidth]{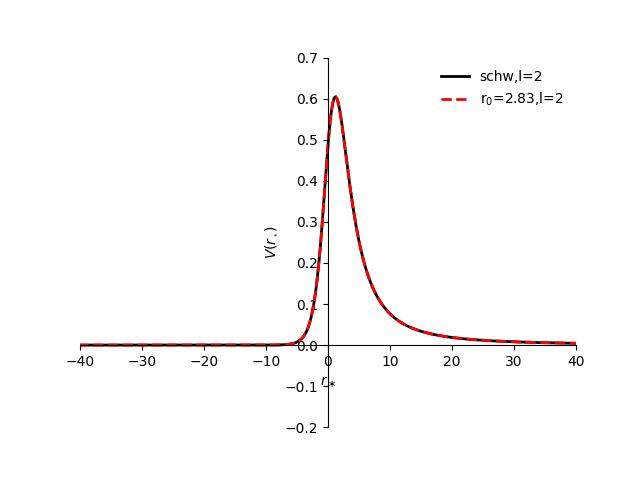}}
		\centerline{(f)}
	\end{minipage}
	\caption{The (e) plot represents the potential function plots for $l$=2, 3 at the gravitational perturbation, with a constant density core parameter of dark matter $r_0$=2.83 kpc. f plot represents the potential function plots for the comparison with the Schwarzschild black hole at the gravitational perturbation, $r_0$=2.83 kpc, $l$=2.}
	\label{fig3}
\end{figure*}

\begin{figure*}[t!]
        \begin{minipage}[t]{0.44\linewidth}
         \centerline{\includegraphics[width=\textwidth]{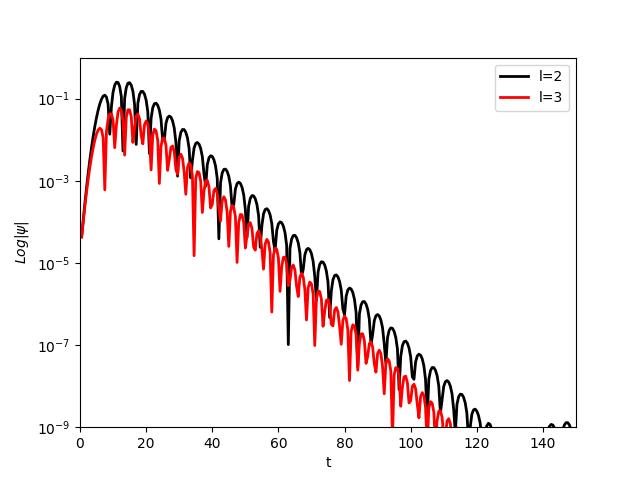}}
		\centerline{(g)}
	\end{minipage}
        \begin{minipage}[t]{0.44\linewidth}
         \centerline{\includegraphics[width=\textwidth]{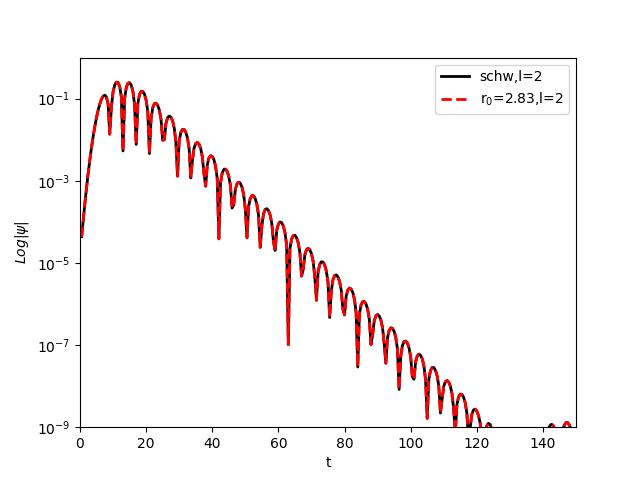}}
		\centerline{(h)}
	\end{minipage}
	\caption{The (g) plot represents the wave function plots for $l$ = 2, 3 at gravitational perturbation, with a constant density core parameter of dark matter, $r_0$ = 2.83 kpc. (h) plot represents the wave function plots for comparison with the Schwarzschild black hole at gravitational perturbation, $r_0$ = 2.83 kpc, $l$ = 2.}
	\label{fig4}
\end{figure*}

\begin{table*}[htbp]
\caption{\label{table2}The quasinormal mode frequencies of gravitational perturbations.}
\begin{tabular}{ccccccc}
\hline
\hline
\multicolumn{1}{c}{$l=2$}\\
\hline
\multicolumn{1}{c}{$r_0$}  &\multicolumn{1}{c}{3th ordel WKB}  &\multicolumn{1}{c}{prony method} \\
\hline
schw &0.7463241297097722-0.17843489444843288$i$ &0.7486789407752416-0.1768265235046571$i$  \\
2.83  &0.746324129709792-0.17843489444852$i$  &0.748678940775244-0.17682652350465133$i$  \\
\hline
\multicolumn{1}{c}{$l=3$}\\
\hline
schw &1.1985302356842915-0.1854568051922124$i$  &1.203757541100846-0.18302901858805987$i$   \\
2.83 &1.1985302356842984-0.18545680519224392$i$ &1.2037575411008923-0.18302901858807674$i$  \\
\hline
\hline
\end{tabular}
\end{table*}

In the gravitational field, the (e) diagram of Fig.\ref{fig3} we draw an image of the potential function in the spacetime of the dark matter constant density core-black hole system for angular quantum numbers $l$ of 2 and 3, which are increasing as the angular quantum number increases.The (f)-diagram in Fig.\ref{fig3} is an image of the potential function at the constant density core parameter of dark matter, $r_0$ = 2.83kpc, and the angular quantum number, $l$ = 2, in comparison with the Schwarzschild black hole.The (g)-map in Fig.\ref{fig4} shows the wavefunction images of the dark matter constant density core-black hole system in spacetime with angular quantum numbers $l$ of 2 and 3, which decay more and more rapidly as the angular quantum number increases.The (h)-map in Fig.\ref{fig4} shows the wavefunction image for the comparison with the Schwarzschild black hole at the constant density core parameter of dark matter, $r_0$ = 2.83, and the angular quantum number, $l$ = 2.

In table\ref{table2} we obtain the frequencies of the spacetime of the constant density dark matter core-black hole system in the gravitational field by applying the third order WKB approximation and prony. From table\ref{table2} we know that the presence of a constant density dark matter core affects the quasinormal modes of the black hole. Most of the errors of the frequencies obtained by applying the prony method and the wkb approximation method are about 0.002, which is in good agreement.

\subsection{Implications for the detectability of the LISA detector}

The formula for the relative deviation of detectability was obtained using the literature\cite{Berti:2005ys,Zhang:2022roh}:
\begin{equation}
2\pi f_{lmn}=Re(\omega_{lmn}),
\end{equation}
\begin{equation}
\tau=-\frac{1}{Im(\omega_{lmn})},
\end{equation}
\begin{equation}
f_{lmn}=f_{lmn}^{Sch}(1+\delta f_{lmn}),
\end{equation}
\begin{equation}
\tau_{lmn}=\tau_{lmn}^{Sch}(1+\delta \tau_{lmn}),
\end{equation}

Here $f_{lmn}$ is the GW frequency, $\tau_{lmn}$ is the damping time of the GW, $\omega_{lmn}$ is the QNM frequency, $f_{lmn}^{Sch}$ is the frequency of the QNM of the schwarzschild black hole, and $\tau_{lmn}^{Sch}$ is the damping time of the QNM of the schwarzschild black hole.

According to the detector\cite{Berti:2005ys},taking the $lmn$= 200 case, using the four equations above, we compute the relative deviation for the angular quantum number $l$=2 under the scalar and gravitational fields as follows

\begin{table}[!htbp]
\centering
\caption{\label{table3}
$\delta f_{lmn}$ (real relative deviation) and $\delta \tau_{lmn}$ (imaginary relative deviation)}
%\begin{ruledtabular}
\begin{tabular}{ccccccc}
\hline
\multicolumn{1}{c}{The scalar field $l=2$}\\
\hline
\multicolumn{1}{c}{$r_0$=2.83kpc}  &\multicolumn{1}{c}{3th ordel WKB}  &\multicolumn{1}{c}{prony method} \\
\hline
$\delta f_{lmn}$     &$3.9968\times10^{-15}$  &$4.21885\times10^{-15}$  \\
$\delta \tau_{lmn}$  &$5.9841\times10^{-14}$  &$9.74776\times10^{-14}$  \\
\hline
\multicolumn{1}{c}{The gravitational field $l=2$}\\
\hline
$\delta f_{lmn}$     &$2.66454\times10^{-14}$  &$3.33067\times10^{-15}$  \\
$\delta \tau_{lmn}$  &$4.88276\times10^{-13}$  &$3.26406\times10^{-14}$  \\
\hline
\end{tabular}
%\end{ruledtabular}
\end{table}

In table 3, we can see that the relative deviation is roughly in the order of $10^{-15}- 10^{-13}$. According to the literature\cite{Berti:2005ys}, the current detection technology cannot reach such a low order of magnitude, but with the development of the detection technology, it is believed that in the future, we may be able to detect the effect of the uniform nucleus of the dark matter on the black holes, which may provide the possibility of detecting the existence of dark matter.

When the constant density dark matter core parameter $r_0=2.83\times10^{-9}$ kpc, i.e., shrinks by a factor of $10^{-9}$ from the original $r_0$. In the gravitational field perturbation when $l=2$, the frequency of qnms is $0.763598-0.138951i$, $0.765560-0.137939i$ by using the third order wkb approximation method and prony method, respectively. the relative deviation is calculated by using the detector, and it is $\delta f_{lmn}=0.0231454$, $\delta \tau_{lmn}=0.284158$, by using the wkb method. And for the prony method is $\delta f_{lmn}=0.0225477$, $\delta \tau_{lmn}=0.281922$. Such relative deviations are much larger than those of the detectors, and in future discovered galaxies, if there are values of constant-density dark-matter core parameters of this magnitude, they will be detected by the detectors.

\section{Summary}
\label{discuss}
In that article, we utilize the quasinormal modes of black holes to detect constant density dark matter cores. By studying the scalar and gravitational field perturbations, we give the potential function image and time evolution image of the black hole containing a constant density dark matter core, and use the third order WKB approximation and prony to obtain the frequency of the QNM. We find that the constant density dark matter core parameter, as one of the parameters of the black hole, affects the quasinormal modes of the black hole under scalar and gravitational field perturbations.

The focus of this paper is as follows:

$\ast$ When the dark matter scale is low, the distribution of dark matter in the galactic core is that of a constant density dark matter core, and we review the spatio temporal metric of the constant density dark matter core-black hole system obtained by Gong et al.

$\ast$ The potential function is obtained using Klein Gordon equation in a scalar field, the potential function image and wave function image are plotted as well as the corresponding QNM frequency is obtained. From Fig.\ref{fig1} we find that as the angular quantum number increases, the peak of the potential function becomes larger and larger. From Fig.\ref{fig2} we know that the ringdown process decays faster and faster as the angular quantum number increases. From table\ref{table1} we know that the presence of a constant density dark matter core affects the quasinormal modes of the black hole. Based on the detectability of the LISA detector, the relative deviation of the gravitational waves was calculated and is shown in table\ref{table3}, where the relative deviation reaches the order of $10^{-15}- 10^{-14}$.

$\ast$ In the gravitational field, the potential function is obtained using the perturbed Ricci tensor and the odd perturbation form of the Regge and Wheeler canonical transformation. The potential function images and wave function images are drawn as well as the corresponding QNM frequencies are obtained. From Fig.\ref{fig3} we find that as the angular quantum number increases, the peak of the potential function becomes larger and larger. From Fig.\ref{fig4} we know that the ringdown process decays faster and faster as the angular quantum number increases. From table\ref{table2} we know that the presence of a constant density dark matter core affects the quasinormal modes of the black hole. And based on the detectability of the LISA detector, the relative deviation of detecting gravitational waves was calculated, which is shown in table\ref{table3} and reaches the order of $10^{-15}- 10^{-13}$.

These studies show that the presence of constant density dark matter cores influences the quasinormal modes of black holes, and we know from table\ref{table3} that the relative deviation reaches $10^{-15}- 10^{-13}$, and if the accuracy of future stochastic gravitational wave background observations is further improved, it will be possible to indirectly detect the existence of constant density dark matter cores near supermassive black holes through stochastic gravitational wave backgrounds, which will offer the possibility of understanding the behavior of dark matter in the vicinity of black holes.

Our current discussion among the scalar and gravitational fields, with other perturbation fields yet to be discussed, as well as the generalization of the constant density dark matter core at the core of galaxies to the case of rotating black holes, are the next focus of our exploration.

\begin{acknowledgments}
We acknowledge the anonymous referee for a constructive report that has significantly improved this paper. We acknowledge the  Special Natural Science Fund of Guizhou University (grant No. X2020068) and the financial support from the China Postdoctoral Science Foundation funded project under grants No. 2019M650846.
\end{acknowledgments}

\end{document}